\documentclass[pra,aps,twocolumn,showpacs]{revtex4}

\usepackage{graphicx}
\usepackage{subfigure}
\begin{document}

\def\be{\begin{equation}}
\def\ee{\end{equation}}
\def\ba{\begin{eqnarray}}
\def\ea{\end{eqnarray}}
\title{Entanglement of a Multi-particle Schr\"odinger Cat State}

\author{A. N. Salgueiro}
\affiliation{Max-Planck Institut f\"ur Quantenoptik, Hans-Kopfermannstr. 1, D-85748  Garching, Germany}


\date{\today}

\begin{abstract}
We characterize the degree of entanglement of a subsystem of $k$ particles in a $N$-two level system  
($k\leq N/2$) initially prepared in a mesoscopic superposition $|\psi\rangle=\int
d\theta f(\theta) (|\phi_{1}(\theta)\rangle^{\otimes N}+|\phi_{2}(\theta)\rangle^{\otimes N})$,
 where $f(\theta)$ is a gaussian or a delta function, subject to the time evolution described by a dephasing channel. 
Negativity is used 
as a measure of entanglement for such system. For an arbitrary number of particles $N$, numerical results are given for the full time evolution up to ten particles. Analytical results are obtained for short times and asymptotic time regimes. 
We show that negativity is initially proportional to the square root of the product of the number of particles in each partition, the overlap ${|\langle\phi_1(\theta)|\phi_2(\theta)\rangle|}^2$ and the coupling to the environment. Asymptotically, negativity tends to zero, a necessary condition for separability.
\end{abstract}

\maketitle

 Generation of macroscopic superpositions has always been a subject of interest, since they
may be viewed as a key to the understanding of some of the striking differences between the quantum and classical descriptions of the world such as the ``nonexistence
at the classical level of the majority of the states allowed by Quantum Mechanics \cite{einstein,sch}.
 Superposition of few photons in a cavity has been demonstrated in Ref.\cite{paris}. Recently, the observation of macroscopic superpositions
  has been proposed for different many particle physical systems,
  such as Bose-Einstein Condensate (BEC)\cite{bec}, superconducting devices (SQUIDs)\cite{squids} and atomic ensembles \cite{ensemble}.
  Quantum Interference in a multi-particle system has been in fact
   observed in superconducting devices (SQUIDs) \cite{squids}. All these phenomena are of relevance for Quantum Computation,  especially because quantum
 computation relies on the creation and maintenance of complex entangled states of many subsystems (superpositions of many particle systems). However, this is a difficult
 task because of decoherence. The interaction between the system and the environment creates entanglement between them and at the same time reduces the entanglement within the system
 itself. An important question in this context is how to quantify and classify entanglement. For example, a way to compare the effective size of a multi-particle macroscopic superposition with the ideal
 GHZ-state (Greenberger-Horne-Zeilinger)  has been shown in Ref. \cite{cat}.

 Pure state entanglement of a bipartite system is presently well understood \cite{pure}. On the other hand,
 entanglement of mixed states
 is still an open problem in spite of all efforts which have been devoted to this problem \cite{mixed}.
 Things get even more
  complicated when one is interested in
investigating the entanglement of multipartite pure or mixed states. However, an important step
 has been given
 in order to determine a computable measure of the entanglement of any mixed state of an arbitrary bipartite system $\{A,B\}$ \cite{vidal}.
  This measure is based on the trace norm of the partial transpose $\rho^{T_A}$ of a bipartite mixed state $\rho$.
  It measures the degree
   to which $\rho^{T_A}$ fails to be positive, and therefore it can be viewed as a quantitative
   version of Peres criterion\cite{peres}. This measure is called negativity and it is defined as ${\mathcal{N}}(\rho)=\sum_i\mid\nu_{i}\mid$
  , where $\nu_{i}$ is the $i$th-negative eigenvalue of the partial transpose $\rho^{T_A}$. In addition,
  negativity is an upper bound to the distillation of entanglement (an asymptotic distillation rate) \cite{vidal}, one of the fundamental measures of
  entanglement.

The aim of the present contribution is to characterize, as a function of time,
the degree of entanglement of a subsystem of 
$k$ particles in an $N$ particle system ($k\leq N/2$), initially prepared in a mesoscopic superposition 

\begin{equation}
\label{um} |\psi\rangle=\int d\theta f(\theta) (|\phi_{1}(\theta)\rangle^{\otimes N}+|\phi_{2}(\theta)\rangle^{\otimes N}),
\end{equation}

\noindent where $|\phi_1(\theta)\rangle=\cos(\theta)|0\rangle+\sin(\theta)|1\rangle$
 and $|\phi_2(\theta)\rangle=\cos(\theta)|0\rangle-\sin(\theta)|1\rangle$ and $f(\theta)=(2\pi s^2)^{-1/2}\exp[-(\theta-\theta_0)^2/(2 s^2)]$ with $s=0$ or $s\neq 0$,
 subject to the time evolution described by the dephasing channel. The dephasing channel is chosen to model decoherence without dissipation. The superoperator which describes the evolution has the form
 has the form
${\mathcal{E}}(\rho)=p_0\rho+(1-p_0)\sigma_z\rho\sigma_z$, where
$p_0=(1+e^{-\gamma t})/2$ and $\sigma_z$ is a Pauli matrix \cite{cat,preskill}. This
quantum channel describes the probability of the qubit to remain
intact or to suffer an "phase flip error". The choice of this
particular evolution is related to the fact that only in this case
an analytical solution for the negativity can be found. We
consider the case where each of the particles is coupled to an
independent environment. In that case, the time evolution of a
system of N particles of two-level systems (qubits) is given by
(see also for example \cite{schu,kempe,cat})

\begin{eqnarray*}
\rho(t)&=&\mathcal{E}(|\phi_1(\theta_0)\rangle\langle\phi_1(\theta_0)|)^{\otimes N}+ \mathcal{E}(|\phi_2(\theta_0)\rangle\langle\phi_2(\theta_0)|)^{\otimes
N}\nonumber\\
&+&\mathcal{E}(|\phi_1(\theta_0)\rangle\langle\phi_2(\theta_0)|)^{\otimes N}+\mathcal{E}(|\phi_2(\theta_0)\rangle\langle\phi_1(\theta_0)|)^{\otimes N}.
\end{eqnarray*}

\textbf{Negativity:}  We calculate the time scale of entanglement of a bipartite ${\{k,N-k\}}$ system of N-particles described by Eq.(\ref{um}) under the time
evolution of the dephasing channel. As a measure of entanglement, we use the negativity $\mathcal{N}(\rho)$. In this case, the negative eigenvalues of the
partial transpose of the density matrix $\rho$ with respect to $k$ subsystem 
have to be evaluated. If the state is separable, the negativity $\mathcal{N}(\rho)=0$. In ref.\cite{adam}, the negativity is used to determine the decay of an upper bound of the entanglement of distillation of two optical qubits interacting in a lossy nonlinear cavity. 
To our knowledge, aside from important mathematical results, there are no explicity numerical or analytical 
results regarding the degree of entanglement of a $N$ particle state
subject to a particular time evolution in the literature in which negativity is used  
as a quantitative measure.

\textbf{$\delta$-function form for $f(\theta)$:}
Let us first consider the simplest case $s=0$, where $f(\theta)$
becomes a delta function. For $t=0$, there are $N-4$ degenerate
null eigenvalues and four non-vanishing eigenvalues of the partial
transpose $\rho^{T_k}$, where one of them is negative, showing
that the system is entangled at $t=0$. These non-vanishing
eigenvalues $\{\lambda_{1,2,3,4}\}$ are given by

\begin{eqnarray}
\lambda_{1,2}&&=\frac{[1\pm\cos^k(2\theta_0)][1\pm\cos^{(N-k)}(2\theta_0)]}{[2 (1+\cos(2\theta_0)^N)]} \nonumber \\
\nonumber \\
\lambda_{3,4}&=&\pm\frac{\sqrt{(1-\cos^{2k}(2\theta_0))}\sqrt{(1-\cos^{2(N-k)}(2\theta_0))}}{[2 (1+\cos(2\theta_0)^N)]} \nonumber
\end{eqnarray}

\noindent and the corresponding eigenvectors ${\{|\lambda_{1,2,3,4}\rangle\}}$  are

\begin{eqnarray*}
&&|\lambda_{1,2}\rangle=|u^{(k)}_{+,-}\rangle\otimes|u^{(N-k)}_{+,-}\rangle \nonumber \\
&&|\lambda_{3,4}\rangle={\frac{1}{\sqrt{2}}}(|u^{(k)}_{+}\rangle\otimes|u^{(N-k)}_{-}\rangle\pm|u^{(k)}_{-}\rangle
\otimes|u^{(N-k)}_{+}\rangle)
\end{eqnarray*}

\noindent with $|u^{(i)}_{\pm}\rangle=\frac{(|\phi_{1}(\theta_0)\rangle^{\otimes (i)}\pm|\phi_{2}(\theta_0)\rangle^{\otimes (i)})}{\sqrt{2[1\pm\cos^i(2\theta_0)]}}$, where $i$ represents the number of particles in each partition, i.e.,  
$i=k$ or $i=N-k$.

\noindent 

The time evolution raises the degeneracy of the $N-4$ null eigenvalues, creating $N-k$ new negative eigenvalues if $N-k>3$. The negative eigenvalues of the
partial transpose $\rho^{T_k}(t)$ can be found for short times ($\gamma t \ll 1$), by solving the equation
$\rho^{T_k}(t)|\Gamma\rangle=\Gamma(t)|\Gamma\rangle$, where $\Gamma(t)$'s are the eigenvalues and
 $|\Gamma\rangle=\sum_{i=1}^4 c_i |\lambda_i\rangle+\sum_{i=1}^4\sum_{k=1}^N d_i^{(k)}\sigma_z^{(k)}|\lambda_i\rangle$ are the corresponding eigenvectors. Some of these new negative eigenvalues
are degenerate. For a general partition $\{k,N-k\}$, there are five negative eigenvalues where two of them are degenerate. From these two negative eigenvalues,
one is $N-k-1$ degenerate and the other one is $k-1$ degenerate. The number of negative eigenvalues is related to the size of the system we consider. We see
thus that in spite of the simplicity of the time evolution (linear and independent coupling to the environment), the appearance of new negative eigenvalues is a clear signature of the considered many particle system .




In order to illustrate this effect, we consider the partition ${\{1,N-1\}}$ 
for which analytical results can be obtained for short
times ($\gamma t\ll 1$).
In this case, negativity is given by

\begin{equation}
\label{uma}
\mathcal{N}(\rho)=-[(N-2)\Gamma_3(t)+\Gamma_2(t)+\Gamma_1(t)]
\end{equation}

\noindent where $\Gamma_1(t)$, $\Gamma_2(t)$ and $\Gamma_3(t)$ are the negative eigenvalues and are given by

\begin{eqnarray}
\label{dois}
\Gamma_1(t)&=&\frac{\lambda_4}{\sqrt{2}} \sqrt{A+\sqrt{A^2-B}} \nonumber \\
\Gamma_2(t)&=&\frac{\lambda_4}{\sqrt{2}} \sqrt{A-\sqrt{A^2-B}} \nonumber  \\
\Gamma_3(t)&=&\frac{\gamma t}{2}\lambda_4 \sqrt{(b_{+}-1)(b_{-}-1)} \nonumber
\end{eqnarray}

\noindent with

\begin{eqnarray}
\label{tres}
A&=&{\big(1-(N+1)\frac{\gamma t}{2}\big)}^2+2(N-1)\frac{\gamma t}{2}[1+ (N+1)\frac{\gamma t}{2}]a_{-}a_{+}\nonumber\\
&&+{\big(\frac{\gamma t}{2}\big)}^2[(N-2)b_{+}+1][(N-2)b_{-}+1]\nonumber \\
B&=&4{\big(\frac{\gamma t}{2}\big)}^2{\big(1-(N+1)\frac{\gamma t}{2}\big)}^2[(N-2)b_{+}-(N-1)a_{-}^2+1]\nonumber\\ &&[(N-2)b_{+}-(N-1) a_{+}^2+1]
\end{eqnarray}

\noindent where

\begin{eqnarray}
a_{\pm}&=&\frac{\cos(2\theta_0)\pm\cos^{(N-2)}(2\theta_0)}{[1\pm\cos^{(N-1)}(2\theta_0)]}\nonumber\\
b_{\pm}&=&\frac{\cos^2(2\theta_0)\pm\cos^{(N-3)}(2\theta_0)}{[1\pm\cos^{(N-1)}(2\theta_0)]}
\end{eqnarray}






\noindent In Fig.~(1) we show the time evolution of the negativity for short time scales: the full curve corresponds to the
(numerical) exact solution  and it coincides with the crosses which show the case where all negative eigenvalues are 
taken into account (given by Eq.(\ref{uma})). The dotted curve corresponds to the evolution of $\Gamma_1(t)$. This figure illustrates 
the importance of taking these new eigenvalues in order to calculate the negativity. In addition it shows the perfect agreement between the numerical exact solution for short times and the analytical calculation of negativity valid for short times, given by Eq.(\ref{uma}). For $N=2$ and $N=3$, there is only one negative eigenvalue which evolves in time and it is the one which already exists at $t=0$, $\lambda_4$. 

\begin{figure}[tbh]
\includegraphics[width=7cm]{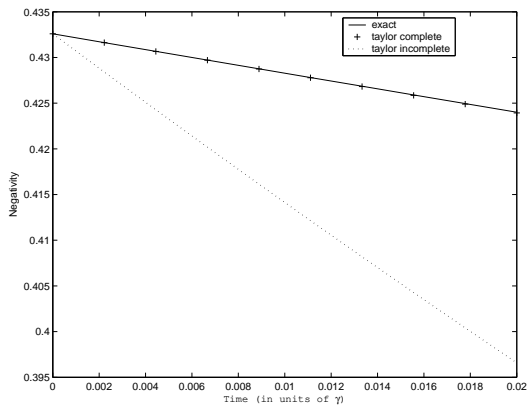}
\caption{Negativity as a function of time for $N=10$, $\theta_0=\pi/3$ and $\gamma=1$. The dotted curve corresponds
to the evolution of $\Gamma_1(t)$. The crosses show the case where all negative eigenvalues are taken into account (Eq.(\ref{uma})) and it coincides with the full curve which corresponds to the numerical exact solution.}
\end{figure}

We are interested in the case of large number of particles $N\gg
1$ where the $\theta_0 \ll 1$.
For the case
$N\theta_0^2\ll 1$,
 one sees that the eigenvalues $\Gamma_2(t)$ and $\Gamma_3(t)$ do not contribute for the negativity. The dominant
 contribution is given by $\Gamma_1(t)$. For the partition $\{1,N-1\}$, the negativity for short times ($\gamma t\ll 1$) and $\theta\ll 1$, such that $N\theta_0\gamma t\ll 1$,
  is given by

\begin{equation}
 \mathcal{N}(\rho)=\sqrt{(N-1)} \theta_0^2 (1-2\gamma t)
\end{equation}

\noindent 
If we consider a general partition $\{k,N-k\}$, one sees that the structure of the problem does not change drastically. 
For this general case in the short time regime ($\gamma t\ll 1$), there are five
negative eigenvalues, where two  of them are degenerate. From these two degenerated negative eigenvalues, one is $N-k-1$ 
degenerate and the other one is $k-1$ degenerate. However, in the case of small $\theta_0$ and $N\gg 1$, 
such that $N\theta_0^2\ll 1$ the main contribution for the negativity is given by
$\Gamma_1(t)$ (where $\Gamma_1(t)=\lambda_4(1+(\gamma t/2)(-N+kA^{(k)}_{+}A^{(k)}_{-}+(N-k)A^{(N-k)}_{+}A^{(N-k)}_{-}))$, with $A^{(i)}_\pm=((\cos{(2\theta_0)}\pm{\cos{(2\theta_0)}}^{(i-1)})/(1\pm{\cos{(2\theta_0)}}^{(i)}))$ where $i$ represents the number of particles in each partition $k$ or $N-k$). In this limit, the negativity is

\[
\mathcal{N}(\rho)=\sqrt{k}\sqrt{(N-k)}\theta_0^2 (1-2\gamma t). 
\]

\noindent This shows that the maximum entanglement is obtained for $k=N/2$, increasing thus the characteristic time scale of the negativity.
In fact even for a more general evolution: depolarizing channel (${\mathcal{E}}(\rho)=p_0\rho+p_1(\sigma_x\rho\sigma_x+\sigma_y\rho\sigma_y+\sigma_z\rho\sigma_z)$, where $p_0=(1+3e^{-\gamma t})/4$, $p_1=(1-e^{-\gamma
t})/4$ and $\sigma_{(x,y,z)}$ are Pauli matrices), we can see numerically that one also gets the same
 dependence for the negativity on the number of particles. This is due to the fact that the
characteristic time scale in this case is essentially governed by initial state correlations.

The full time evolution of the entanglement of the subsystem under investigation is given by 
numerical calculations up to ten particles for an arbitrary $\theta_0$. Fig.~2 illustrates 
the behavior of the negativity for $N=10$ and $\theta_0=\pi/3$ as a function of time (numerical exact calculation).

\begin{figure}[thb]
\includegraphics[width=7cm]{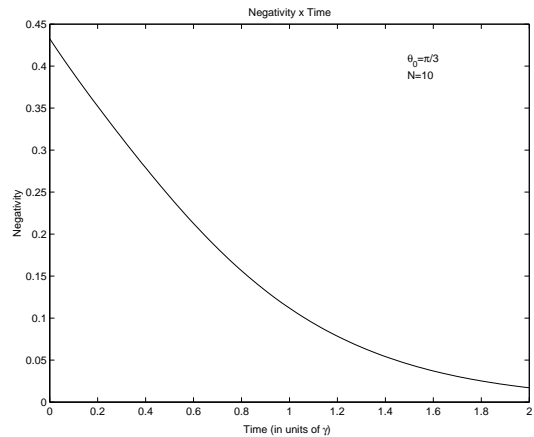}
\caption{The full curve represents the exact numerical solution for the time evolution of the negativity for all times.
For short times, the behavior of this curve can be seen in Fig.~1 and it coincides with the analytical solution of 
negativity given by Eq.(\ref{uma}). In the asymptotic limit, the negativity goes to zero. In this figure, we consider $\{N-1,1\}$ 
partition, $\theta_0=\pi/3$ and $N=10$.}
\end{figure}

In the asymptotic limit, it is also possible to calculate the negativity and it is zero.
This is a necessary condition for the separability of the system under investigation.

The particular case $\theta_0=\pi/4$ in Eq.(\ref{um}) (GHZ-state
of the type $|\psi\rangle=|0\rangle_x^{\otimes
N}+|1\rangle_x^{\otimes N}$) and the evolution under the dephasing
channel, the negativity becomes independent of $N$. One recovers
the $N$-dependence if $|0\rangle_x \rightarrow |0\rangle_z$ and
$|1\rangle_x \rightarrow |1\rangle_z$. These particular situations
illustrate the dependence of the time evolution on the initial
condition: If the evolution operator is "orthogonal" (in the sense
just described) to the initial condition, one slows down the
entanglement process. Also if the initial state of Eq.(\ref{um})
has $|\phi_1\rangle=|0\rangle$ and
$|\phi_2\rangle=\cos(\theta_0)|0\rangle+\sin(\theta_0)|1\rangle$,
the negativity is completely governed by the time evolution of the negative eigenvalue which exists for $t=0$. 
In this case, the state $|\phi_1\rangle\langle|\phi_1|$ is an eigenstate of the evolution operator and we
observe that the degeneracy among the $N-4$ null initially
existing eigenvalues is not removed.

\textbf{Gaussian form for $f(\theta)$:} Next we investigate the general situation where a finite width $s$ in
$f(\theta_0)$ appears in Eq.(\ref{um}). The gaussian form does not affect the structure of the negative eigenvalues.
For a general partition $\{k,N-k\}$ and short time regime ($\gamma t\ll 1$), there are five negative eigenvalues, 
where two of them are degenerate. 
From these two negative eigenvalues one is $N-k-1$ degenerate and the other one $k-1$ degenerate. For the partition
$\{1,N-1\}$, the general form of the eigenvalues of Eq.(\ref{dois}) remains unchanged, only the coefficients 
$a_{\pm}$ and $b_{\pm}$ are more complicated. Analytical expressions can be obtained for $\sigma^2\ll 1$ and for short times $\gamma t\ll 1$. Their explicit form will not be given here, since they are not particularly illuminating. We see from the numerical implementation of the analytical result that the negativity has the same dependence in the number of particles as in the previous case ($s=0$). As can be seen from Fig.~3 the present situation introduces a slight modification of the previous result. Fig.~3 shows the negativity as a function of the number of particles $N$ for a $\{N-1,1\}$ partition 
when $f(\theta)$ is a gaussian or delta function. Fig.~3 was generated from the analytical expressions of negativity 
for short times in both cases (delta and gaussian).

\begin{figure}[thb]
\includegraphics[width=7cm]{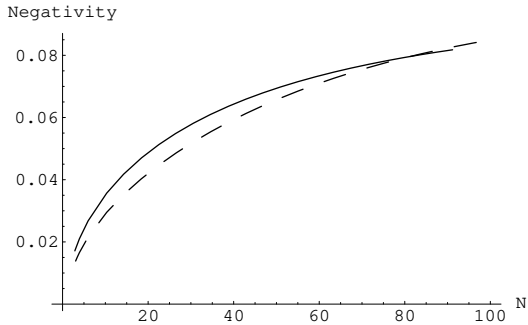}
\caption{Negativity as a function of the number of particles $N$,
for fixed $\theta_0=0.1$, $\gamma t=0.02$, for $\{N-1,1\}$
partition. The dashed curve represents the negativity as a function of the number of particles for the delta case ($s=0$) 
while the full curve shows the negativity as function of the number of particles for the gaussian case with $s=0.05$.}
\end{figure}

In summary, we have determined 
the degree of entanglement of a subsystem of $k$ particles in this $N$ particle ($k\leq N/2$) system. 
We observed that negativity is proportional to the square root of the number of particles in each partition, 
the overlap between the states involved in the initial mesoscopic superposition and the coupling to the environment
, in the limit of small angle and short times.  In addition, the value obtained here for the negativity gives an upper bound for the distillation rate, since it has
been shown in ref. \cite{vidal} the relation between these two quantities.
It would be interesting to investigate other dynamical situations, in particular, where the evolution can be
derived from first principles. The case of a competition between collective and independent particle-environment coupling also deserves further investigation in order to see how
universal are the conclusions we came to.

\begin{acknowledgments}
ANS acknowledges J. I. Cirac for stimulating discussions and  for suggesting the problem, M. Wolf, J.J. Garcia Ripoll, M. C. Nemes, A. F. R. de Toledo Piza, H. Weidenm\"uller and M. Weidem\"uller for valuable discussions.
\end{acknowledgments}


\begin{thebibliography}{99}

\bibitem{einstein} Letter from Albert Einstein To Max Born in 1954, cited by E. Joos in {\it New Techniqes
and Ideas in Quantum Measurement Theory}, edited by D. M. Greenberger (New York Academy of Science, New York, 1986).

\bibitem{sch}  E. Schr\"odinger, Die Naturwissenschaften {\bf 23}, 807 (1935).

\bibitem{paris} M. Brune {\it et al.}, Phys. Rev. Lett. {\bf 77}, 4887 (1996).

\bibitem{bec} J. I. Cirac, M. Lewenstein, K. Molmer and P. Zoller, Phys. Rev. {\bf A 57}, 1208 (1998);
J. Ruostekoski, M. J. Collect, R. Graham and D. F. Walls, Phys. Rev. {\bf A 57}, 511 (1998).

\bibitem{squids}  J. R. Friedman {\it et al.}, Nature {\bf 406}, 43
(2000); C. H. van der Wal \emph{et al.}, Science \textbf{290}, 773 (2000).

\bibitem{ensemble}  S. Massar and E. S. Polzik, Phys. Rev. Lett. {\bf 91}, 060401 (2003).

\bibitem{cat} W. D\"ur, C. Simon and J. I. Cirac, Phys. Rev. Lett. {\bf 89}, 210402 (2002).

\bibitem{pure} C. H. Bennett, H. J. Bernstein, S. Popescu and B. Schumacher, Phys. Rev. {\bf A 53}, 2046 (1996).

\bibitem{mixed} C. H. Bennett, D. P. DiVicenzo, J. A. Smolin and W. K. Wootters, Phys. Rev. {\bf A 54}, 3824 (1996);
P. M. Haydern, M. Horodecki and B. Terhal, quant-ph/0008134; V. Vedral and M. Plenio, Phys. Rev. {\bf A 57}, 1619 (1998); L. Henderson and V. Vedral, Phys.
Rev. Lett. {\bf 84}, 2263 (2000); M. Horodecki, P. Horodecki and R. Horodecki, Phys. Rev. Lett. {\bf 84}, 2014 (2000); W. K. Wootters, Phys. Rev. Lett. {\bf
80}, 2245 (1998); G. Vidal, Phys. Rev. {\bf A 62}, 062315 (2000).

\bibitem{vidal} G. Vidal and R. F. Werner, Phys. Rev. {\bf A 65}, 032314 (2002).

\bibitem{peres} A. Peres, Phys. Rev. Lett. {\bf 77}, 1413 (1996).




\bibitem{preskill} J. Preskill, "Lectures Notes for Physics 229: Quantum Information and Quantum Computation" (1998).

 \bibitem{schu} B. Schumacher, Phys. Rev. {\bf A 54}, 2614 (1996).

 \bibitem{kempe} C. Simon and J. Kempe, Phys. Rev. {\bf 65}, 052327 (2002).

\bibitem{adam} Adam Miranowicz, quant-ph/0402025.









(2001).





\end{thebibliography}
\end{document}